\documentclass[aps,prfluids,superscriptaddress]{revtex4-2}
\usepackage{graphicx}
\usepackage{dcolumn}
\usepackage{bm}

\usepackage{xcolor}
\usepackage{caption}
\usepackage{subcaption}

\begin{document}

\preprint{APS/123-QED}

\title{Delaying Leading Edge Vortex Detachment by Plasma Flow Control at Topologically Critical Locations}
\author{Johannes Kissing}
\email{kissing@sla.tu-darmstadt.de}
\author{Bastian Stumpf}
\affiliation{Technische Universität Darmstadt, Fluid Mechanics and Aerodynamics (SLA)\\
Alarich-Weiss-Straße 10, D-64287 Darmstadt, Germany}
\author{Jochen Kriegseis}
\affiliation{Karlsruhe Institute of Technology (KIT), Institute of Fluid Mechanics (ISTM)\\
Kaiserstr. 10, D-76131 Karlsruhe, Germany}
\author{Jeanette Hussong}
\author{Cameron Tropea}
\affiliation{Technische Universität Darmstadt, Fluid Mechanics and Aerodynamics (SLA)\\
Alarich-Weiss-Straße 10, D-64287 Darmstadt, Germany}

\date{\today}

\begin{abstract}
Flapping wing propulsion offers unrivalled manoeuvrability and efficiency at low flight speeds and in hover. These advantages are attributed to the leading edge vortex developing on an unsteady wing, which induces  additional lift. We propose and validate a manipulation hypothesis that allows  prolongation of the leading edge vortex growth phase, by delaying its detachment with the aid of flow control. This approach targets  an overall lift increase on unsteady airfoils. A dielectric barrier discharge plasma actuator is successfully used to compress secondary structures upstream of the main vortex on a pitching and plunging flat plate. To determine flow control timing and location, the tangential velocity on the airfoil surface is used, which is also used to quantify topological effects of flow control. This flow control is then tested for different motion kinematics and on a NACA 0012 airfoil. Significant increase of the peak circulation of the leading edge vortex of about 20\,\% for all cases with flow control indicates that this approach is applicable for various kinematics, dynamics and airfoil types.
\end{abstract}

\maketitle

\textit{Introduction.} Biological propulsion based on flapping wings offers unique advantages in lift and manoeuvrability at low flight Reynolds numbers. Both features are essential to increase the performance and efficiency of micro-air vehicles (MAV) and energy harvesting devices or to enhance their gust tolerance.\\
Flapping flight aerodynamics are dominated by the leading edge vortex (LEV), which grows on a wing when the effective inflow angle changes rapidly. The LEV forms by a roll-up of the separated leading edge shear layer, caused by an adverse pressure gradient at the leading edge. The emergence of the LEV and its effects on aerodynamic forces on the airfoil constitute the dynamic stall phenomena, as discussed by \cite{CARR.1988}. Formation and subsequent growth of the LEV induces increased transient lift. This induced lift drops when the vortex detaches, which is accompanied by a high pitching moment. It is known that insects gain higher transient lift at low Reynolds numbers with an attached LEV \cite[cf.][]{Dickinson.1999,Ellington.1996} and that this higher lift is prolonged with the aid of active feedback control to keep the vortex attached to their wings \cite{Ristroph.2010}.\\
On the one hand, LEV induced forces can be beneficial for applications where higher lift is desired. On the other hand, aerodynamic forces can be detrimental with respect to the structural integrity of helicopter or wind turbine blades due to the induced pitching moment during vortex detachment.
When aiming to exploit beneficial LEV effects further for MAVs or energy harvesting devices,  manipulation of the LEV is a frequently considered approach \cite[cf.][]{Eldredge.2019}. Numerous studies attempt to  trap vortices on steady airfoils with the aid of passive or active flow manipulation. Potential flow analysis of a trapped vortex on a steady airfoil using a leading edge flap and suction by \cite{Rossow.1978} indicated an order of magnitude higher lift potential (also denoted as super lift). The experimental feasibility of this approach was demonstrated by \cite{Riddle.1999}.\\
Until now flow control studies on unsteady airfoils focused on a reattachment of the leading edge shear layer in terms of separation control. Here, the net lift enhancement is achieved by reducing the effect of lift decrease after LEV detachment. Permanent separation control inevitably suppresses the LEV formation and thus high transient lift, which is reasonable if the vortex induced effects are intended to be mitigated to maintain structural integrity.\\
A multitude of  flow control actuators have been used successfully for separation control. Examples include synthetic jets \cite{Taylor.2015}, suction and blowing devices \cite{Greenblatt.2001} and plasma actuators \cite{Post.2006}. Plasma actuators have the advantage of being fast acting and very lightweight due to the absence of moving parts. Both features are especially important when considering highly dynamic wing motion and resulting inertial forces.\\
Closed-loop separation control with a plasma actuator at the leading edge of a pitching NACA 0012 airfoil was first tested by \cite{Lombardi.2013}. By adding momentum to the leading edge shear layer they achieved a faster reattachment of the shear layer during pitch up. The highest net lift increase was achieved when separation control was temporally disabled, such that a LEV could form. With the aid of pulsed plasma actuators on the leading edges of vertical axis wind turbine (VAWT) blades, \cite{BenHarav.2016} were able to increase the turbine net power by ten \%. They conclude that this is also achieved by reattaching the leading edge shear layer more rapidly, and additionally, by an LEV that is closer to the blade surface.\\
In the present study, a flow control approach to exploit the LEV induced lift is introduced and experimentally validated. This approach adapts the biological principle of an enhanced LEV induced lift instead of suppressing the vortex by separation control, and was first proposed by \cite{Kutemeier.2019}. More specifically, higher lift is assumed to be achieved by a prolongation of the LEV growth phase, which  in turn is achieved by delaying the vortex detachment process. The body force induced by a plasma actuator applied at topological critical locations will be utilized to achieve the prolonged LEV growth phase.\\ 
After the introduction of the manipulation hypothesis by means of topological arguments, the implementation on a pitching and plunging flat plate will be demonstrated by means of time-resolved particle image velocimetry (PIV) flow field measurements. Topological measures and effects on LEV characteristics will serve to test the effectiveness of the hypothesis in detail. Finally, the robustness of the proposed flow control concept will also be tested with different motions kinematics of the flat plate and complementary tests on a NACA 0012 airfoil.

\begin{figure}
\captionsetup[subfigure]{position=top,singlelinecheck=off,justification=raggedright}
    \begin{subfigure}[t]{0.475\textwidth}
    \caption{}
        \includegraphics[width =\columnwidth, trim = 0 0 0 0, clip]{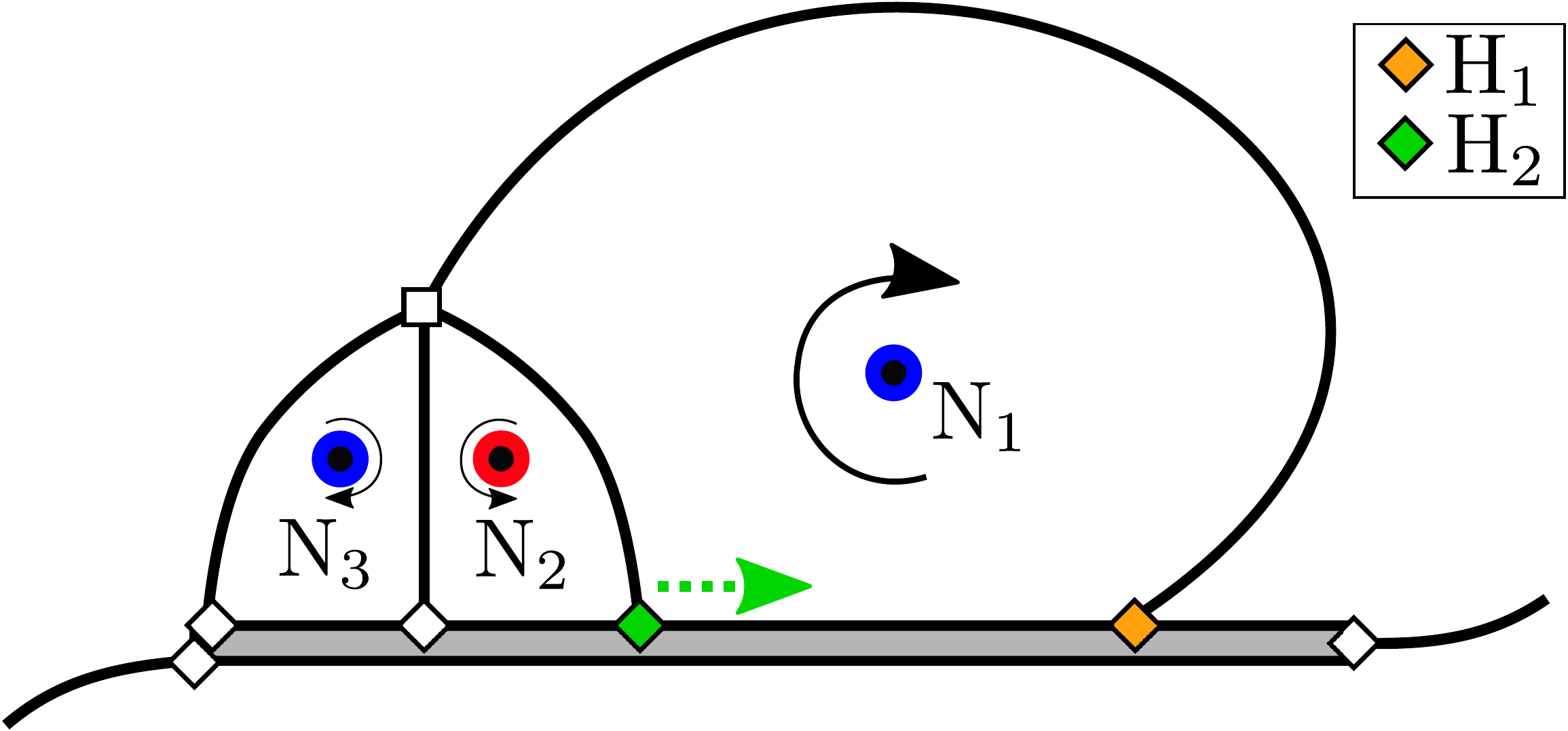}
        \label{fig:Detach_BLE}
    \end{subfigure}
    \begin{subfigure}[t]{0.475\textwidth}
    \caption{}
        \includegraphics[width =\columnwidth, trim = 0 0 0 0, clip]{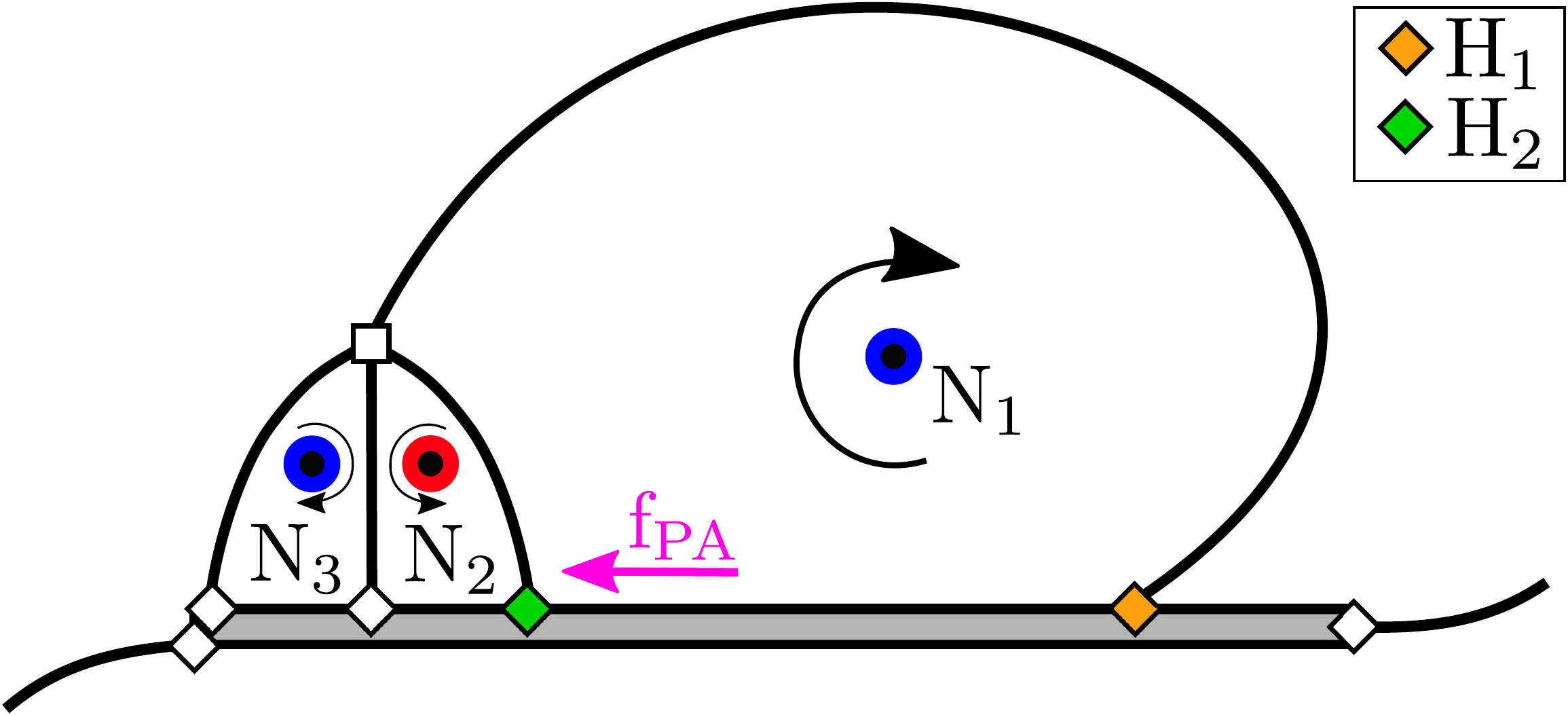}
        \label{fig:ManHyp}
    \end{subfigure}
\caption{\label{fig:Topo} Topological sketch of the flow field around a pitching and plunging flat plate with a leading edge vortex (LEV), denoted as node N\textsubscript{1}, growing on the airfoil. Nodes are indicated by circles, a full saddle by a square and half saddles by diamonds. (a) LEV detachment caused by growing secondary vortices  (also termed secondary structures and shown as nodes N\textsubscript{2} and N\textsubscript{3}), in accordance with the boundary layer eruption mechanism. (b) Manipulation hypothesis to delay LEV detachment by compression of secondary vortices with the plasma induced body force f\textsubscript{PA}.}
\end{figure}

\begin{figure}[b]
\begin{subfigure}[t]{0.455\textwidth}
    \includegraphics[width =\textwidth, trim = 0 5 90 15, clip]{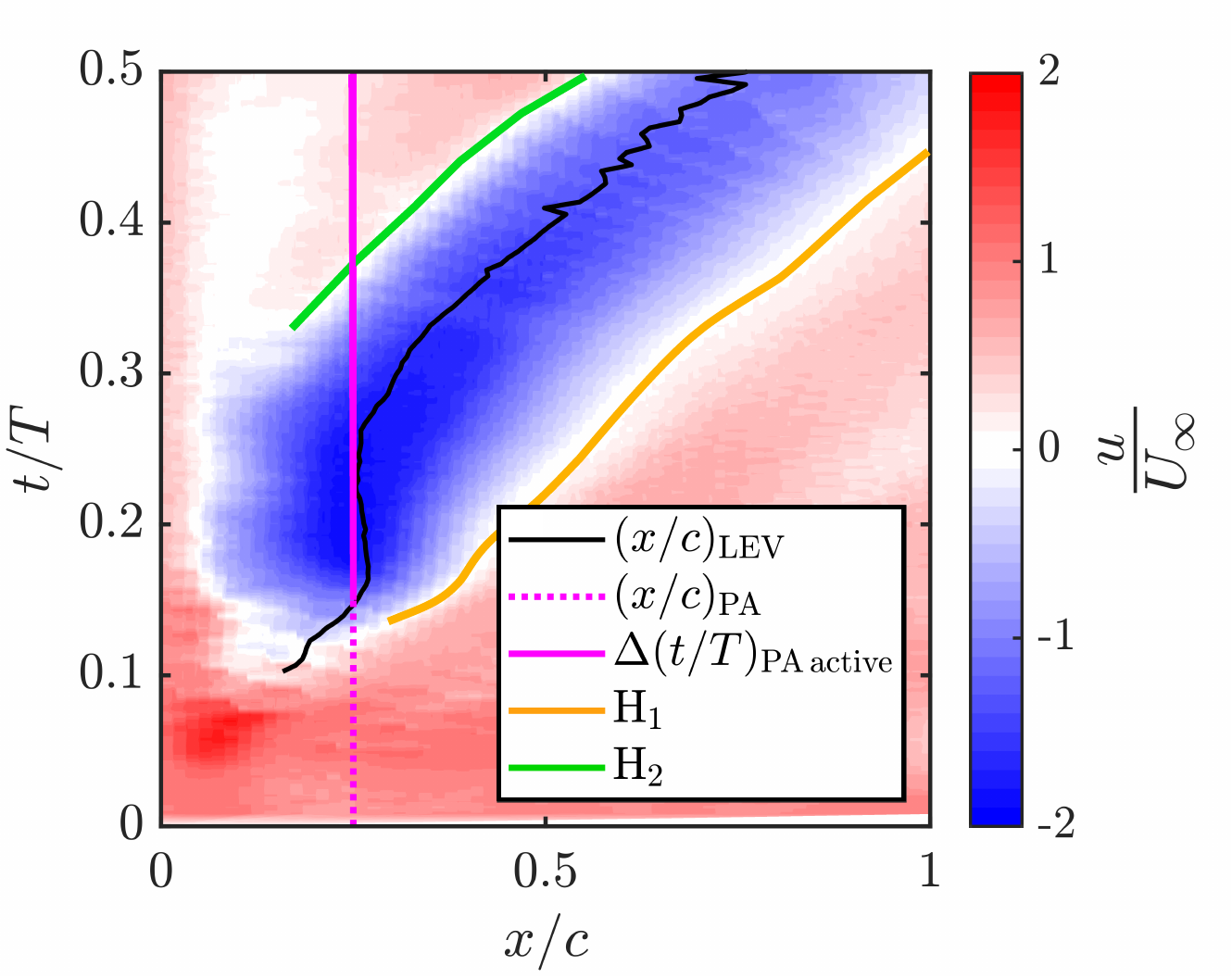}
    \caption{Base}
    \label{fig:RecircBase}
\end{subfigure}
\begin{subfigure}[t]{0.51\textwidth}
    \includegraphics[width =\textwidth, trim = 45 5 10 15, clip]{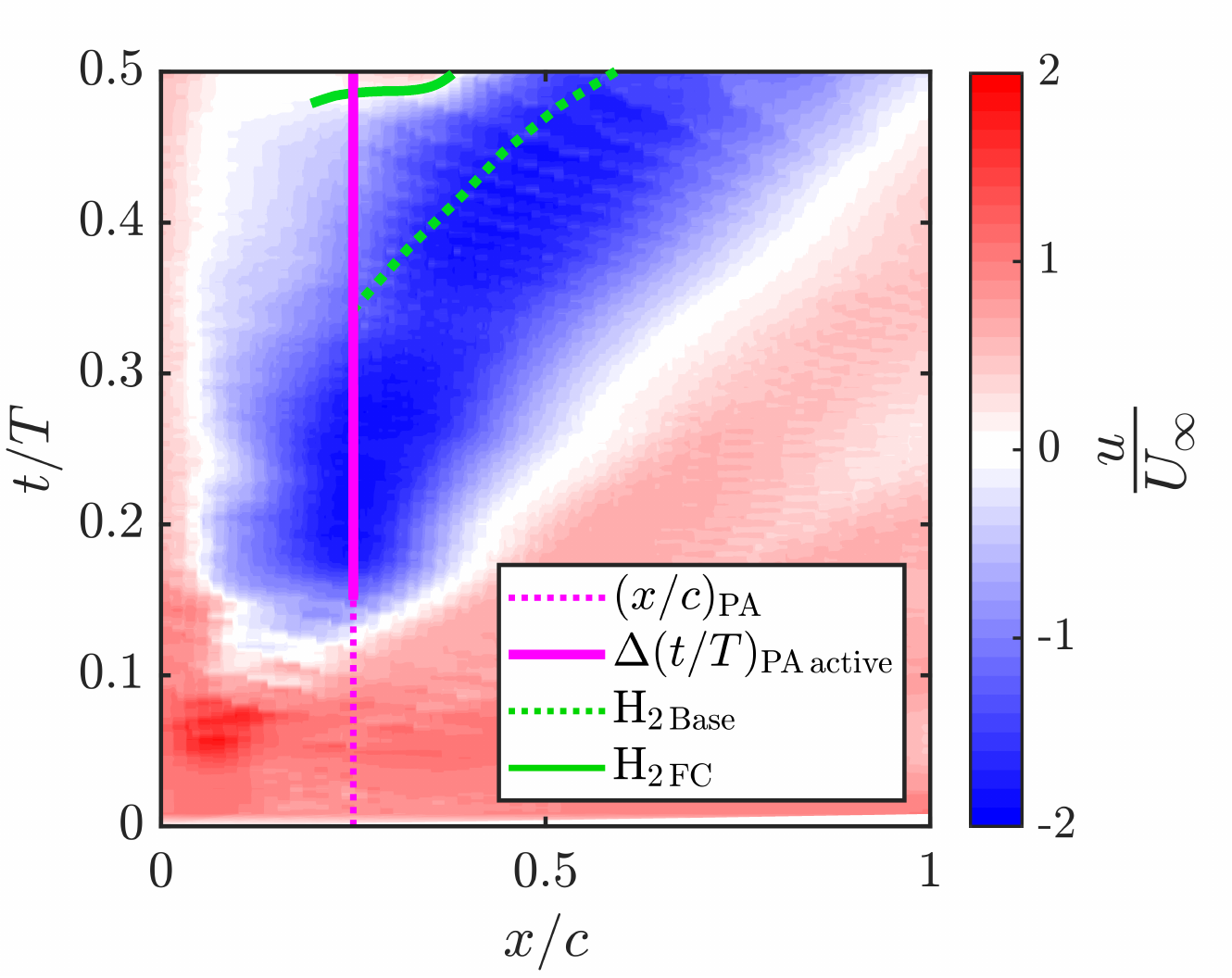}
    \caption{FC}  
    \label{fig:RecircPA}
\end{subfigure}
\caption{Tangential velocity $u$ on the airfoil surface over dimensionless time $t/T$, shown in a plate fixed frame of reference over the dimensionless chord position $x/c$ with the origin at the leading edge ($x/c=$ 0). (a) Evolution without flow control. The leading edge vortex center position $(x/c)_{\mathrm{LEV}}$ and LEV confining half-saddles (H\textsubscript{1} and H\textsubscript{2}) are indicated. The location of the plasma actuator $(x/c)_{\mathrm{PA}}$ and the actuation period $\Delta(t/T)_{\mathrm{PA\,active}}$ are also indicated. (b) Evolution with flow control applied. The secondary structure confining half saddle in the uncontrolled (H\textsubscript{2 Base}) and the controlled case (H\textsubscript{2 FC}) are compared, to evaluate flow control effects.}
\end{figure}

\textit{Manipulation hypothesis.} The proposed manipulation hypothesis, which targets a prolonged LEV growth phase to attain higher net lift, is based on the suppression of the LEV detachment process and follows the approach introduced in \cite{Kutemeier.2019}. For nominally two-dimensional airfoils in flapping flight, two vortex detachment mechanisms are known: A detachment triggered by viscous effects and denoted as boundary-layer eruption \cite{doligalski1994vortex,Widmann.2015} and the so-called bluff body detachment \cite{Rival.2014}, with the chord of the airfoil as a characteristic length scale. Both mechanisms can be explained with a topological flow field analysis as introduced in \cite{Foss.2004} and adapted to the dynamic stall phenomenon by \cite{Rival.2014}. This concept is based on a constant Poincar\'{e} index, computed from nodes, which are vortex centers, saddles, which are saddle points in the flow and half saddles, which are stagnation points on the airfoil.\\
For a boundary layer type LEV detachment, the growth of secondary vortices, which arise from a viscous response of the boundary layer on the airfoil upstream of the main vortex, initiates the detachment. This can be illustrated by considering the flow topology during LEV growth, depicted in Fig.~\ref{fig:Detach_BLE}. 
Secondary vortices, which are denoted secondary structures, are denoted here as nodes N\textsubscript{2} and N\textsubscript{3}. Once they grow, their rear confining half saddle (highlighted as a green diamond and denoted as H\textsubscript{2} in Fig.~\ref{fig:Detach_BLE}) moves towards the trailing edge. As soon as H\textsubscript{2} merges with the LEV confining half saddle (highlighted as orange diamond and denoted as H\textsubscript{1} in Fig.~\ref{fig:Detach_BLE}), the LEV detaches.\\
To compress the secondary structures - thus delay LEV detachment - plasma actuators are applied and operated in upstream direction on the plate so as to counter-act the H\textsubscript{2}-advection towards the trailing edge. The compression is realized with the aid of the plasma induced body force f\textsubscript{PA}, as depicted in Fig.~\ref{fig:ManHyp}.\\
Determination of actuation location and timing requires an identification of topologically key characteristics in terms of half saddles on the airfoil surface. As outlined by \cite{Rival.2014}, the tangential velocity immediately above the airfoil is a good indicator of half saddles, through its change of velocity sign. An exemplary evolution of the tangential velocity on the flat plate airfoil in the uncontrolled baseline case is shown in Fig.~\ref{fig:RecircBase}. 
At $t/T=$ 0 the airfoil is at the initial top position and at 0.5 at its final bottom end position.\\
The diagonal blue trace in Fig.~\ref{fig:RecircBase} indicates the upstream orientated (negative) tangential velocity induced by the LEV while it convects over the airfoil. The rear confining half saddle of the LEV (H\textsubscript{1} in Fig.~\ref{fig:Detach_BLE}) can be identified by the change of velocity sign downstream of the LEV, as indicated by the orange line in Fig.~\ref{fig:RecircBase}. In the upstream direction the LEV is separated from secondary structures by another half saddle (H\textsubscript{2} in Fig.~\ref{fig:Detach_BLE}) and thus another change of velocity sign, marked in Fig.~\ref{fig:RecircBase} with a green line. In the uncontrolled case in Fig.~\ref{fig:RecircBase}, H\textsubscript{2} can be identified from about $t/T=$ 0.325 by the change of velocity sign upstream of the LEV.\\
From a topological point of view, successful compression of secondary structures would result in a  half saddle located further upstream of H\textsubscript{2} (green line in Fig.~\ref{fig:RecircBase} to the top left).\\
To test whether secondary structures can be compressed using the plasma induced body force f\textsubscript{PA}, and whether that leads to a prolonged growth phase of the LEV, the plasma actuator is placed at $x/c=$ 0.25 and switched on from $t/T=$ 0.15 to 0.5 for motion parameters of the baseline case. The plasma actuator location is chosen under consideration of the trade-off between an upstream orientated location for an early actuation and a downstream orientation for a longer effective compression of secondary structures before they grow beyond the plasma actuator. Compared to the instant when H\textsubscript{2} can be identified in the uncontrolled case ($t/T=$ 0.325), the plasma body force is deployed very early to make sure that secondary structures are compressed, even before they are visible in the tangential velocity plot.

\textit{Setup and data processing.} Flow control experiments are conducted on airfoils that execute combined pitching and plunging motion kinematics in an open return wind tunnel at Technische Universit\"at Darmstadt. The turbulence level in the test section of 0.45\,m\,$\times$\,0.45\,m width and height was measured by hot-wire anemometry and found to be below 2\% for $U_{\infty}=$ 3.45\,m/s.\\
Both investigated airfoil types, a flat plate and a NACA 0012, have a chord  of $c=$ 120\,mm. The flat plate airfoil has an asymmetric leading edge of 30\,$^{\circ}$ edge angle, resulting in a distinct shear layer separation.\\ 
Actuators used for flow control in this study are dielectric barrier discharge plasma actuators (DBD-PA), which consist of five layers of Kapton$^{\copyright}$ tape of 325\,$\mu$m overall thickness as a dielectric medium. The alternating current high voltage is generated with a \textit{Minipuls 2.1} generator. The plasma actuator is operated at 10.8\,kV with a carrier frequency of $\varphi=$10\,kHz, where the maximum induced velocity by the plasma body force is about 5.43\,m/s according to \cite{Kriegseis.2013}. Laser light reflections from the DBD-PA are reduced by an additional layer of black paint on the dielectric surface.\\
The airfoils are mounted on a mid-span bracket attached to two linear actuators of type \textit{LinMot PS01 - 48x240F - C}, one at the leading edge and another at about 68 \% of the chord. Combined pitching and plunging kinematics are realized by execution of different motion profiles on both actuators. To ensure accurate and jitter-free motion, both actuators are  monitored online by position and acceleration sensors.\\
Effective inflow conditions on the airfoil can be described by a combination of dimensionless numbers and the effective inflow angle on the leading edge of the airfoil $\alpha_{\mathrm{eff}}$. The Reynolds number $Re=c U_{\infty}/\nu$, which determines the horizontal inflow velocity, is computed using the chord $c$, free stream velocity $U_{\infty}$ and the kinematic viscosity $\nu$. The reduced frequency $k=\pi f c/U_{\infty}$, with the motion frequency $f$, and the Strouhal number $St=2 \hat{h} f/U_{\infty}$, with the plunging amplitude $\hat{h}$, determine the vertical inflow velocity on the airfoil via the plunging velocity $\dot{h}$. Finally, $\alpha_{\mathrm{eff}}$ results from the superposition of the induced angle of attack due to the vertical plunging motion of the airfoil $\alpha_{\mathrm{plunge}}= \dot{h}/U_{\infty}$ and the geometric angle of attack $\alpha_{\mathrm{geo}}$ determined by the pitching angle. To produce a distinct effective inflow angle amplitude $\hat{\alpha}_{\mathrm{eff}}$, $\alpha_{\mathrm{geo}}$ is set according to $\alpha_{\mathrm{plunge}}$, which is in turn determined by $\dot{h}$ via $St$ and $k$. All investigated kinematics in this study consider a quasi-sinusoidal effective inflow angle evolution, and only the downstroke of the airfoil.\\
Time-resolved 2D2C particle image velocimetry (PIV) is used to characterize the manipulation effects on the flow field and LEV. The laser light sheet is produced by a \textit{Litron} LDY - 303 PIV high-speed Nd:YLF laser with 527 nm wavelength and light sheet optics centred at quarter span. A narrow band-pass filter (530\,nm$\pm$10\,nm) in front of the camera lens is used to eliminate any discharge light emitted by the plasma ($<$430\,nm). DEHS seeding particles with a mean diameter of 1.0\,$\mu$m and a response time $\tau_{s}$ = 2.7\,$\mu$s according to \cite{Raffel} are introduced into the settling chamber of the wind tunnel. Image pairs are recorded at 1\,kHz with a pulse delay of $\Delta t$ =150\,$\mu$s. Commercial software (\textit{PIVview2C}) is used for standard FFT cross-correlation with a Gaussian sub-pixel peak detection. A multi-grid, multi-pass interrogation scheme of 64\,px initial and 16\,px final square IA size at 50\% overlap is used for correlations. The spatial resolution of 251\,$\mu$m/px leads to 59 velocity information per chord. 2.3\% velocity outliers have been identified by a normalized median test of threshold 2 \cite[cf.][]{Westerweel.2005}, which have been replaced by a 3x3 neighborhood interpolation.\\
Effects of flow control on LEV circulation and center position are evaluated by means of a vortex identification methodology as proposed by \cite{Graftieaux.2001}. This methodology uses two scalar fields ($\Gamma_{1}$ and $\Gamma_{2}$) to identify the vortex center position and vortex boundary separately via thresholding. To reduce the influence of fluctuations on the standard deviation of the circulation, each velocity component is temporally filtered using a second order Savitzky-Golay filter of $\Delta t$ = 11\,ms length (11 velocity information instants), as introduced by \cite{Rival.2014}.\\
Ensemble averaged LEV circulation and vortex center position for each number of repetitions (1 to 20) are randomly sampled to determine the convergence of the standard deviation with the number of repetitions $i$. A total number of 1000 bootstrap samples are taken for each value of $i$ to estimate the population standard deviation \citep[cf.][]{Efron.1979}. 15 repetitions yield reasonable convergence of the standard deviation within $\pm$ 5\% of the asymptotic value. Thus, all cases are recorded 15 times.

\begin{figure*}
\begin{subfigure}[t]{0.237\textwidth}
        \includegraphics[width=\textwidth, trim = 10 5 130 10, clip]{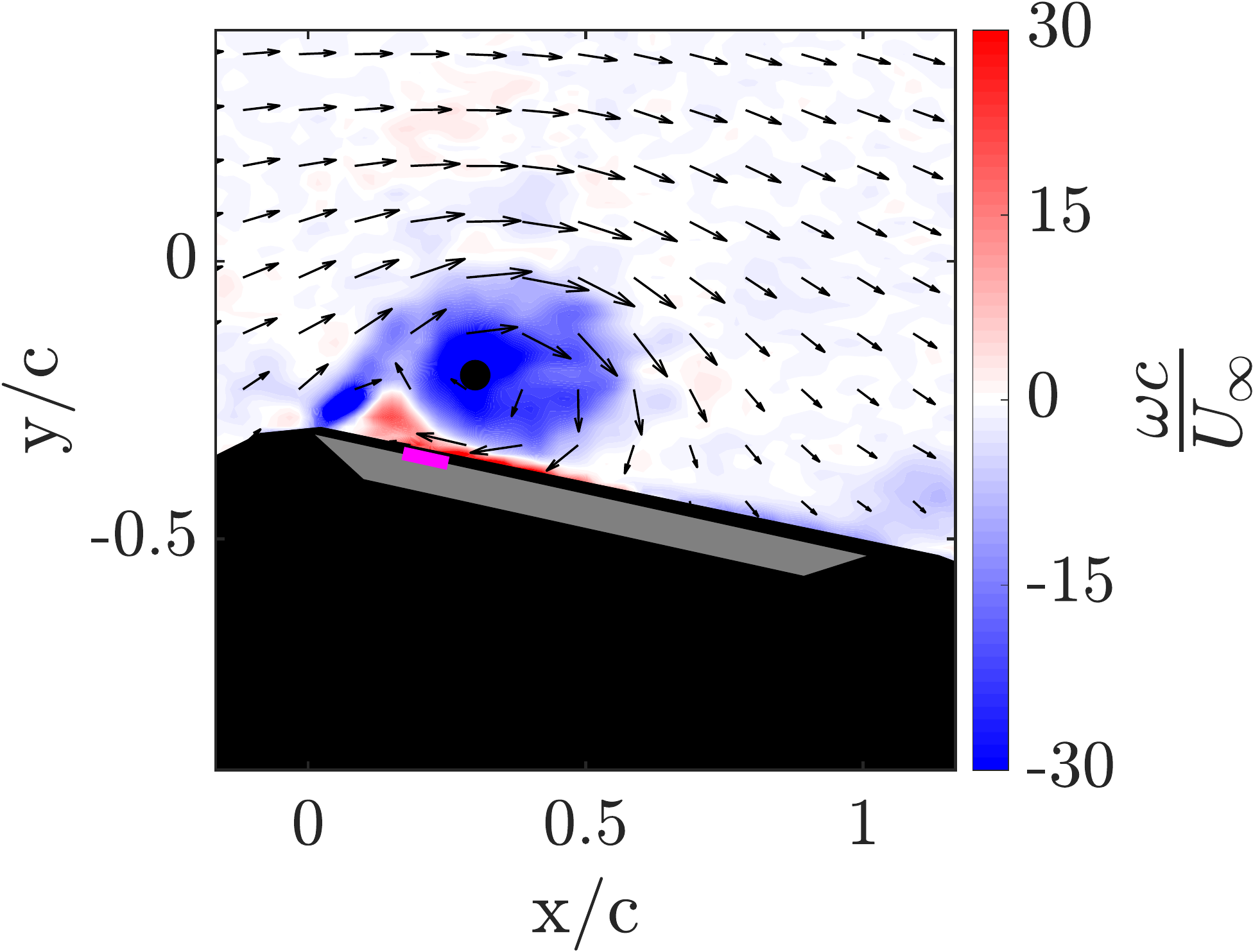}
        \caption{Base; $t/T$ = 0.25}
        \label{fig:Vort_Base_tT0x25}
        \end{subfigure}
        \begin{subfigure}[t]{0.189\textwidth}
        \includegraphics[width=\textwidth, trim = 95 5 130 10, clip]{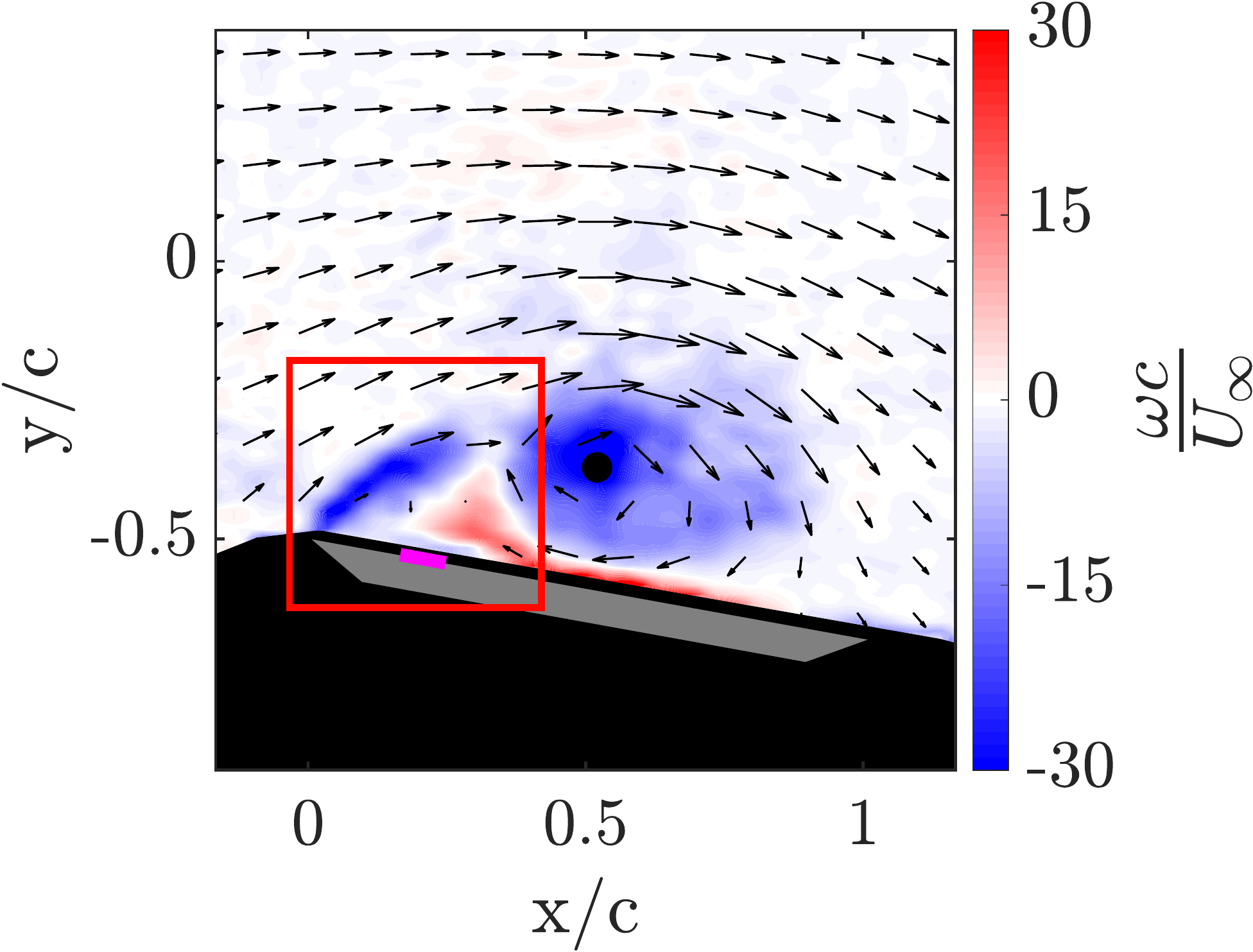}
        \caption{Base; $t/T$ = 0.35}
        \label{fig:Vort_Base_tT0x35}
        \end{subfigure}
        \begin{subfigure}[t]{0.189\textwidth}
        \includegraphics[width=\textwidth, trim = 95 5 130 10, clip]{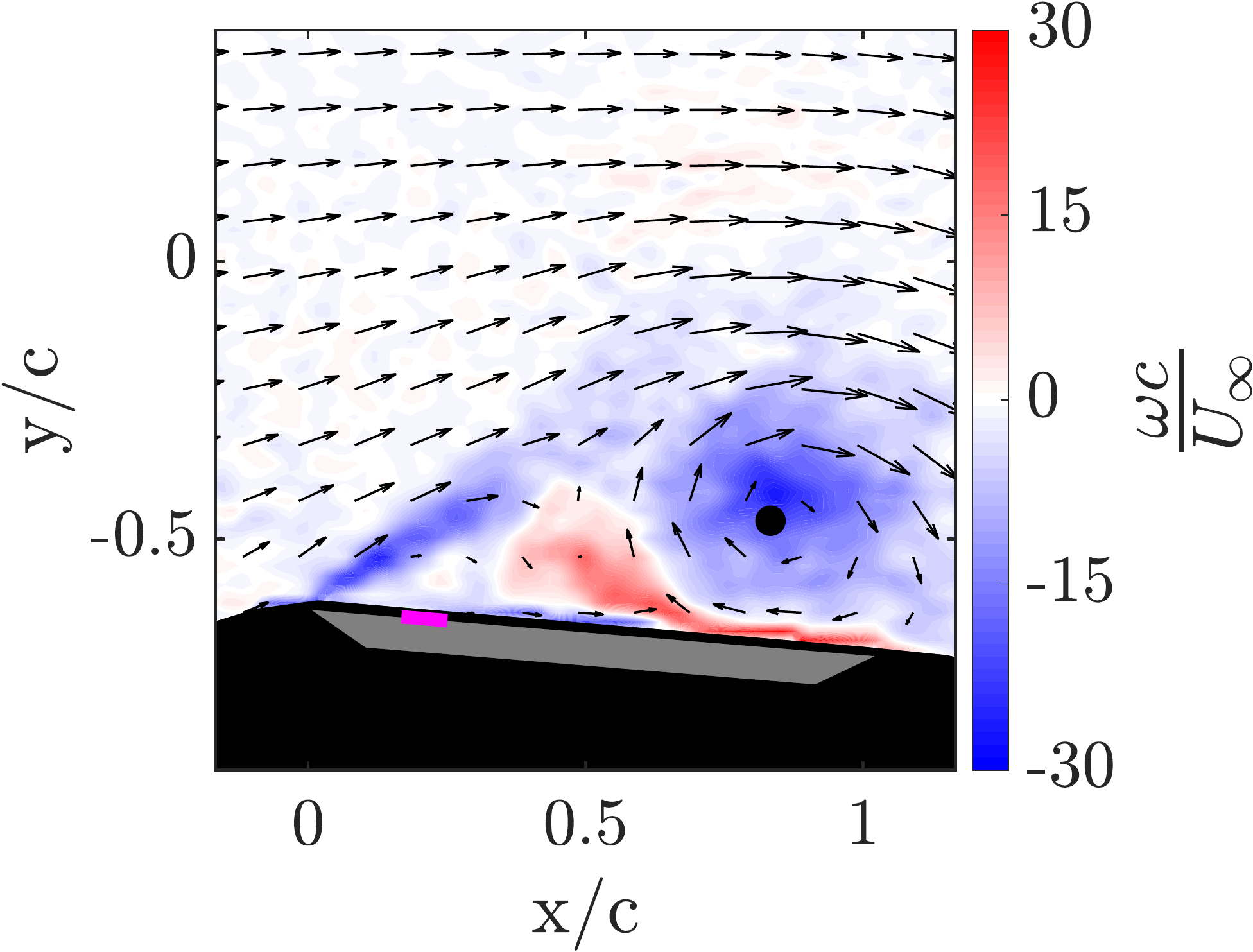}
        \caption{Base; $t/T$ = 0.45}
        \label{fig:Vort_Base_tT0x45}
        \end{subfigure}
        \begin{subfigure}[t]{0.315\textwidth}
        \includegraphics[width=\textwidth, trim = 0 0 0 0, clip]{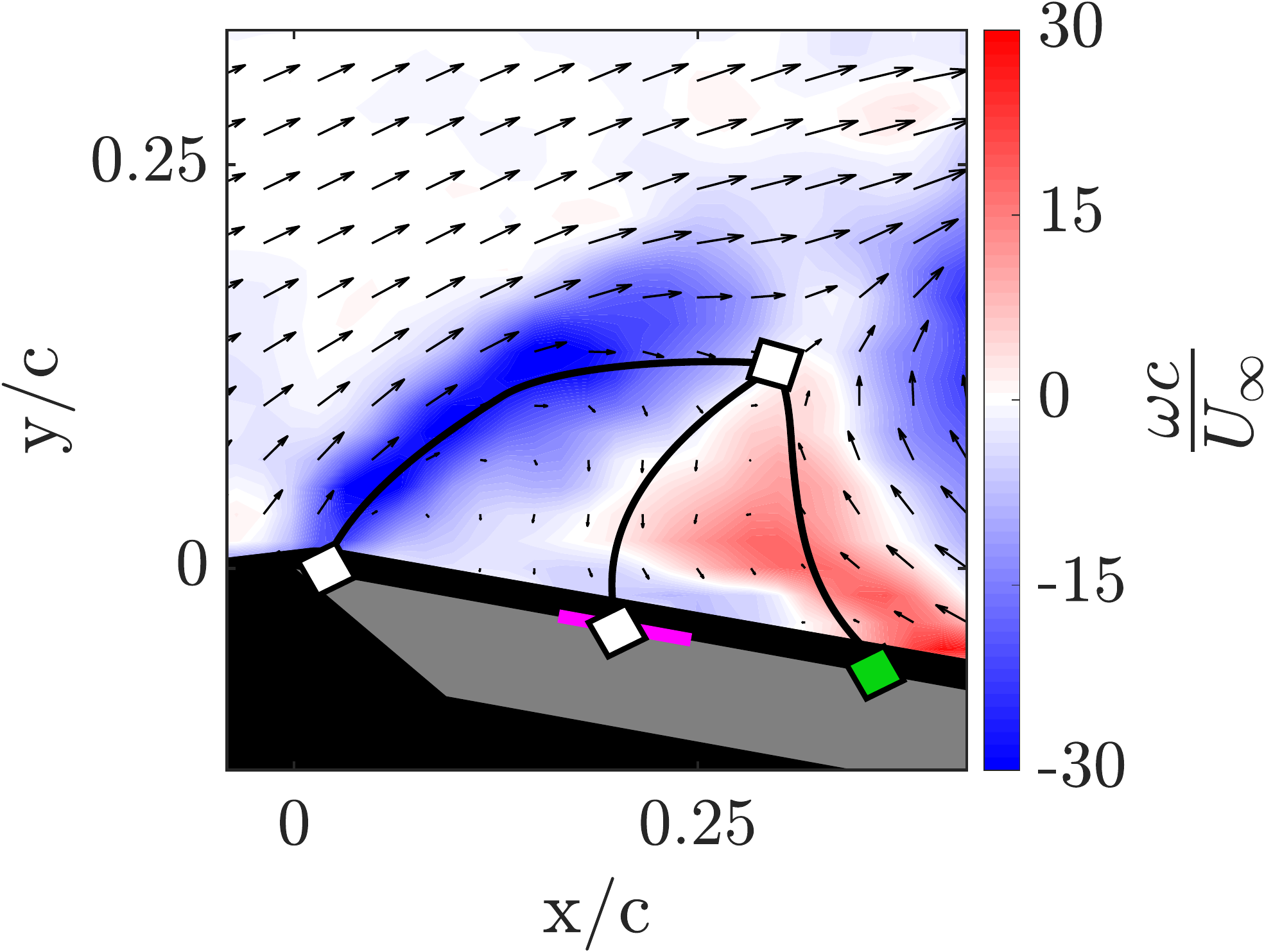}
        \caption{Base; $t/T$ = 0.35}
        \label{fig:Vort_Base_tT0x45_Zoom}
        \end{subfigure}
	    \begin{subfigure}[t]{0.237\textwidth}
        \includegraphics[width=\textwidth, trim = 10 5 130 10, clip]{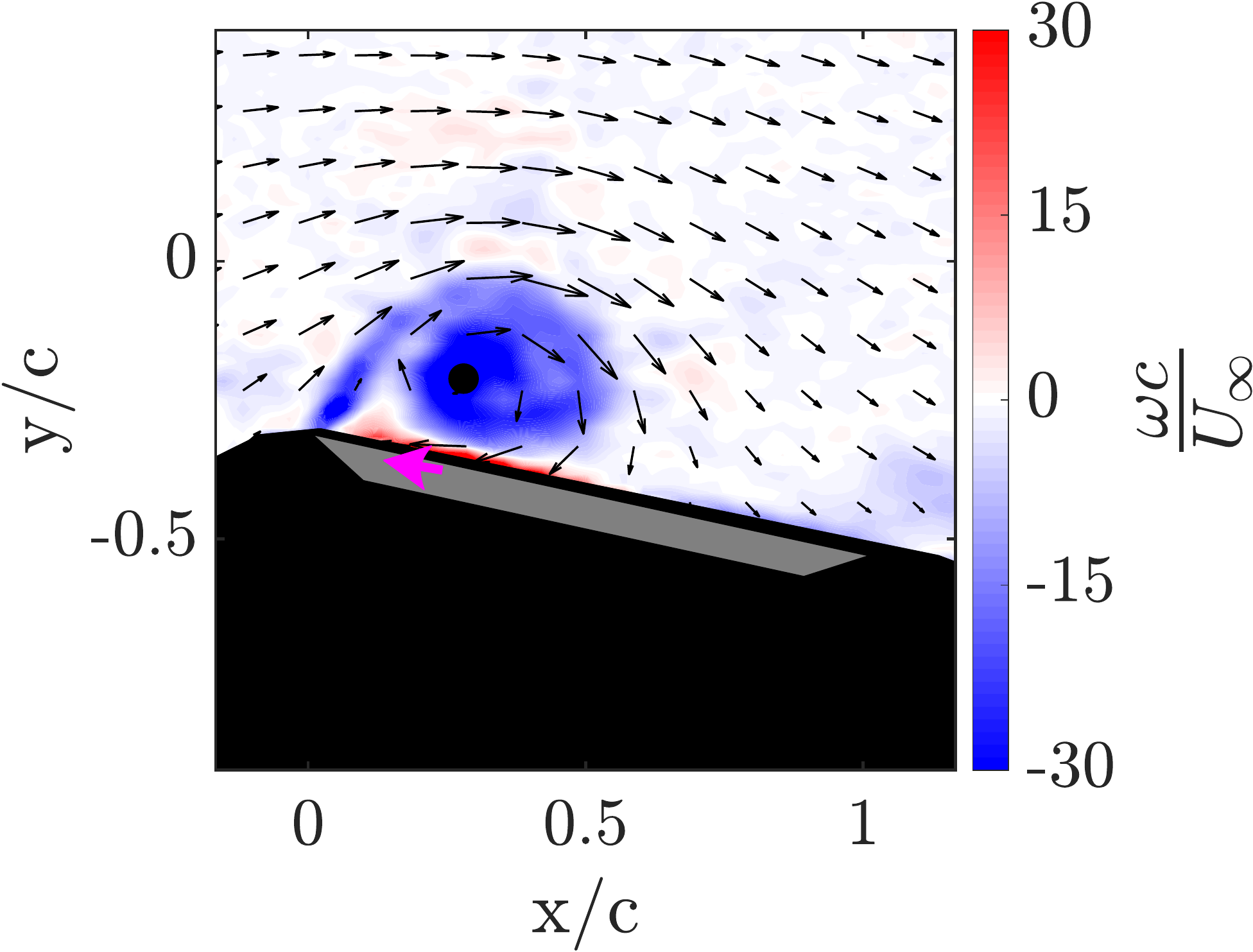}
        \caption{FC; $t/T$ = 0.25}
        \label{fig:Vort_FC_tT0x25}
        \end{subfigure}
        \begin{subfigure}[t]{0.189\textwidth}
        \includegraphics[width=\textwidth, trim = 95 5 130 10, clip]{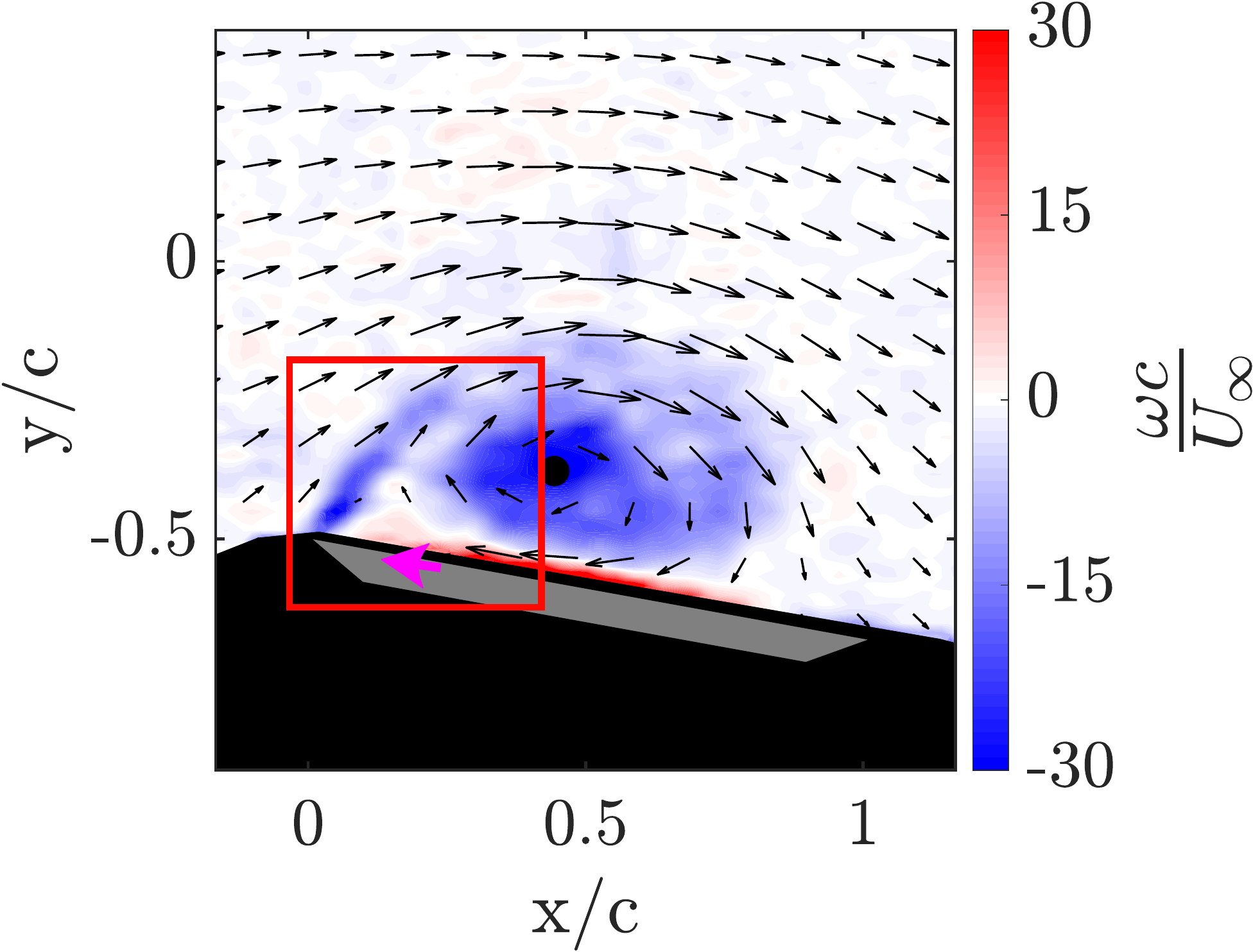}
        \caption{FC; $t/T$ = 0.35}
        \label{fig:Vort_FC_tT0x35}
        \end{subfigure}
        \begin{subfigure}[t]{0.189\textwidth}
        \includegraphics[width=\textwidth, trim = 95 5 130 10, clip]{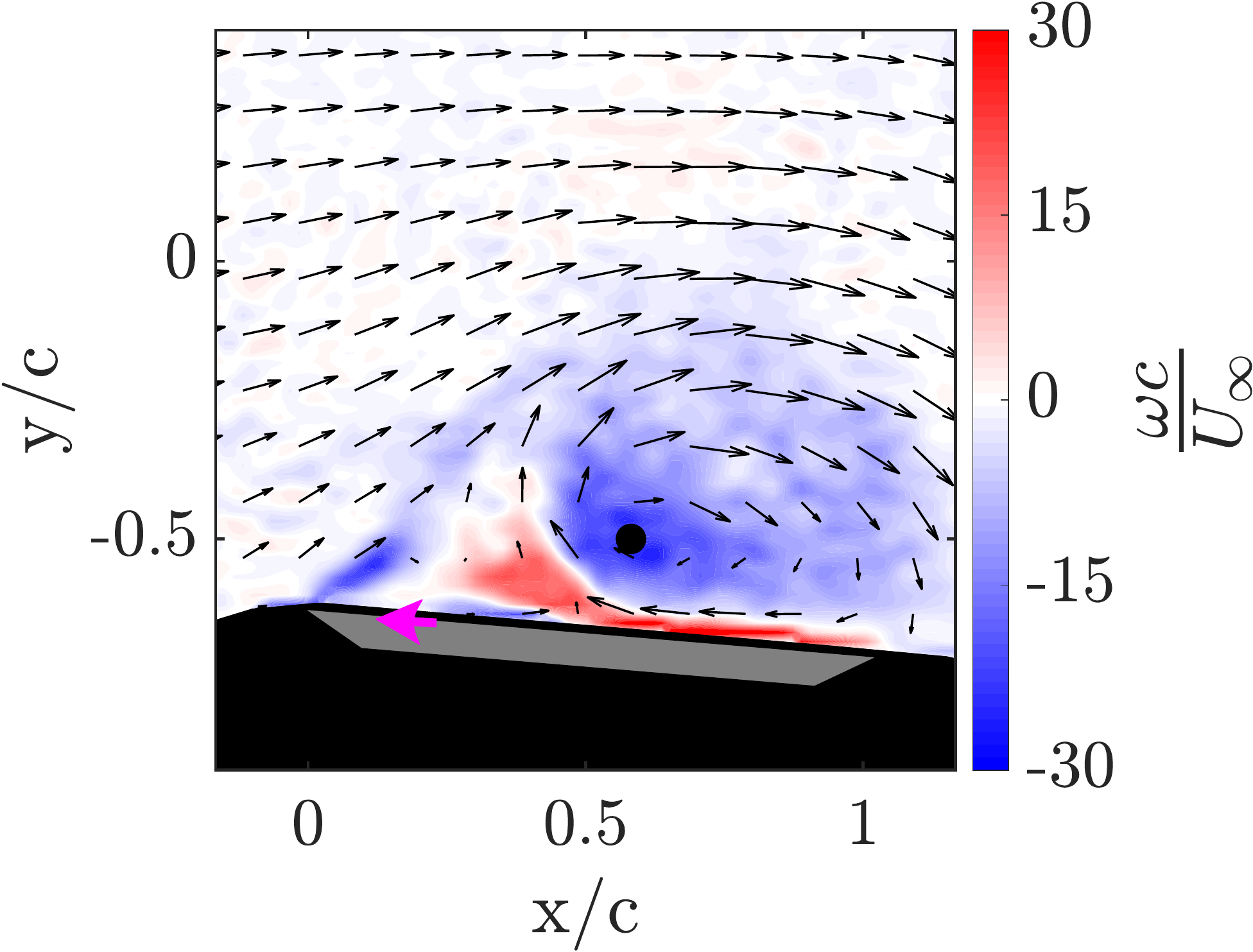}
        \caption{FC; $t/T$ = 0.45}
        \label{fig:Vort_FC_tT0x45}
        \end{subfigure}
        \begin{subfigure}[t]{0.315\textwidth}
        \includegraphics[width=\textwidth, trim = 0 0 0 0, clip]{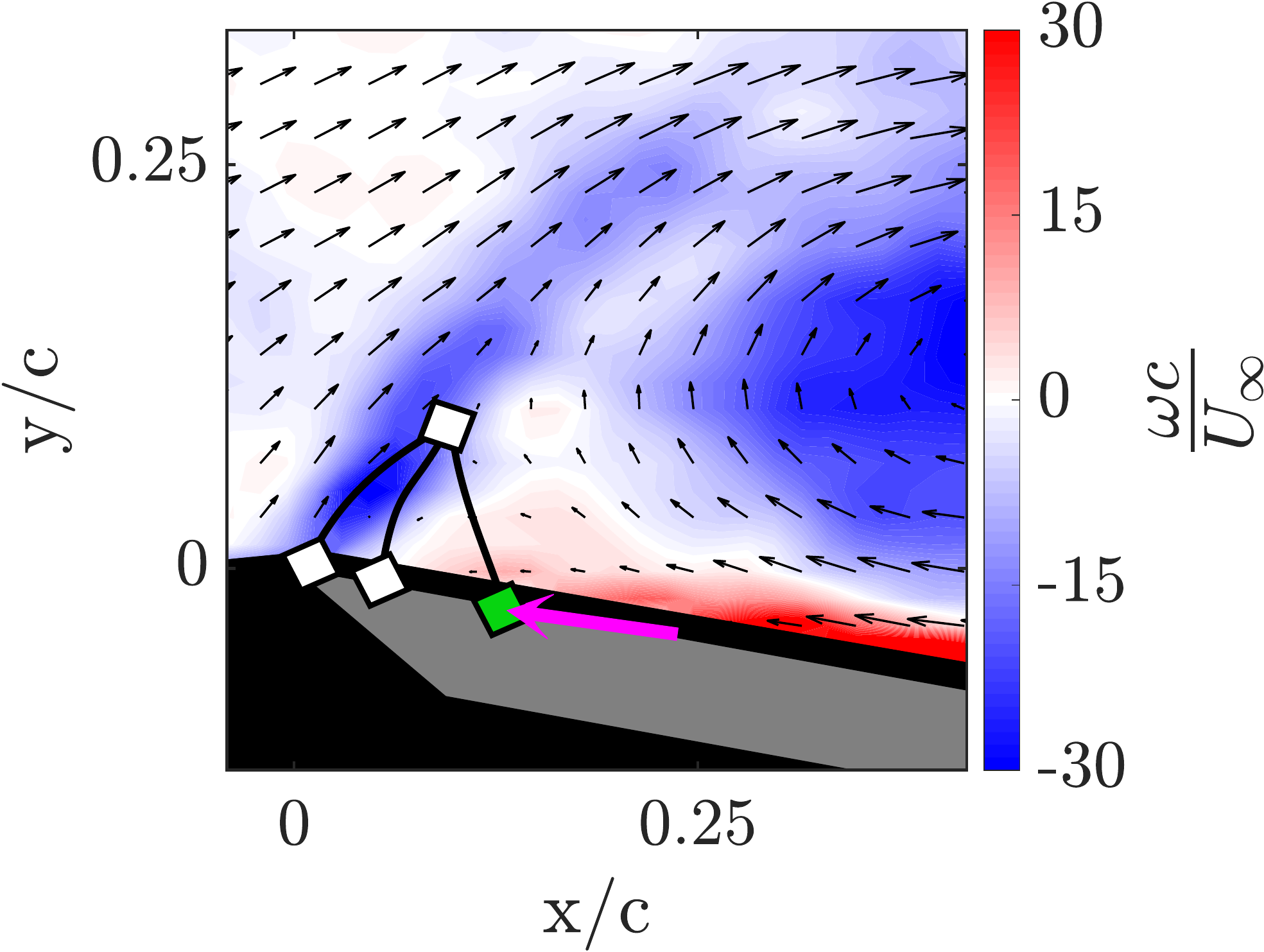}
        \caption{FC; $t/T$ = 0.35}
        \label{fig:Vort_FC_tT0x45_Zoom}
        \end{subfigure}
\caption{\label{fig:FlowFields} Velocity field and normalized vorticity $\omega c/ U_{\infty}$ for different dimensionless time instants $t/T$. (a)-(d) without flow control (Base); (e)-(h) controlled case (FC). An inactive plasma actuator is indicate by a magenta line and an active one by a magenta arrow. While (a)-(c) and (e)-(g) depict the entire field with only every sixth velocity vector for clarity, (d) and (h) show the leading edge region with every second vector as indicated by a red square in (b) and (f). The inflow is from the left, the airfoil is masked out in grey and the laser light shadow caused by the airfoil in black.}
\end{figure*}

\textit{Manipulation effects on flow fields.} A first detailed evaluation of the effectiveness of the manipulation hypothesis is conducted on a pitching and plunging flat plate airfoil. The free stream velocity $U_{\infty}$ is set such that $Re$ yields 22,000. The airfoil executes a one-shot downstroke motion at a reduced frequency $k$ of 0.48 and a Strouhal number $St=$ 0.1, at $\hat{\alpha}_{\mathrm{eff}}=$ 30$\,^{\circ}$. These parameters are referred to as the baseline case  in the following.\\
To investigate flow control effects qualitatively, flow fields for the baseline case without flow control (Base) and with flow control (FC) are contrasted in Fig.~\ref{fig:FlowFields}.
Videos of both flow fields in direct comparison are available online (\url{http://dx.doi.org/10.25534/tudatalib-175}). The vortex center, obtained from the $\Gamma_{1}$ scalar field, is  marked by a black circle. The LEV can be identified as the clockwise rotating region of negatively signed (blue color-coded) vortical fluid. It grows by accumulation of vorticity from the leading edge shear layer.\\
For both cases, a small region of positively signed (red color-coded) vorticity upstream of the LEV is evident already in the first depicted dimensionless time instant $t/T=$ 0.25 in Fig.~\ref{fig:Vort_Base_tT0x25} and Fig.~\ref{fig:Vort_FC_tT0x25}. This region is indicative for the secondary vortex N\textsubscript{2}.\\
For subsequent time instants in the uncontrolled case, secondary structures grow and the connection between the LEV and the leading edge shear layer becomes weaker (e.g. at $t/T=$ 0.35 in Fig.~\ref{fig:Vort_Base_tT0x35}). This process is accompanied by large secondary structures, which are topologically identified in the zoomed leading edge flow field in Fig.~\ref{fig:Vort_Base_tT0x45_Zoom}. The LEV finally detaches from the leading edge shear layer at about $t/T=$ 0.45 in Fig.~\ref{fig:Vort_Base_tT0x45}.\\
In contrast, secondary structures are hardly identifiable in the controlled case at $t/T=$ 0.35 in Fig.~\ref{fig:Vort_FC_tT0x35}. This is also topologically evaluated in Fig.~\ref{fig:Vort_FC_tT0x45_Zoom}, where very small secondary structures can be observed in comparison to the case without flow control. The LEV center is more upstream in the controlled case when the plasma is deployed. These observations indicate a successful compression of secondary structures ahead of the main LEV by the plasma actuator and, consequently, a reduced vortex center convection.\\
To confirm secondary structure suppression, the footprint of the topological structures is shown in Fig.~\ref{fig:RecircPA} for immediate comparison with the baseline case (Fig.~\ref{fig:RecircBase}). To evaluate flow control effects, the rear confining half saddle of secondary structures, H\textsubscript{2}, is compared with and without flow manipulation. To enable comparability, the trajectory of H\textsubscript{2} in the uncontrolled case is projected into Fig.~\ref{fig:RecircPA}, where it is marked by a dashed green line.\\
In comparison to the case without flow manipulation, where H\textsubscript{2} was identifiable from $t/T=$ 0.325 onward, it arises delayed at about 0.47 in the case when flow control is deployed. For later instants during  the downstroke, where both half saddles can be compared directly, H\textsubscript{2 FC} is located closer to the leading edge, indicated by lower $x/c$ values.\\
The delayed emergence of the rear confining half saddle of secondary structures, and its position closer to the leading edge of the airfoil, confirms a successful compression of secondary structures, as intended in the manipulation hypothesis and illustrated in Fig.~\ref{fig:ManHyp}.

\begin{figure}[b]
\captionsetup[subfigure]{position=top,singlelinecheck=off,justification=raggedright}
\begin{subfigure}[t]{0.475\textwidth}
    \caption{}
    \includegraphics[width =\textwidth, trim = 0 5 25 15, clip]{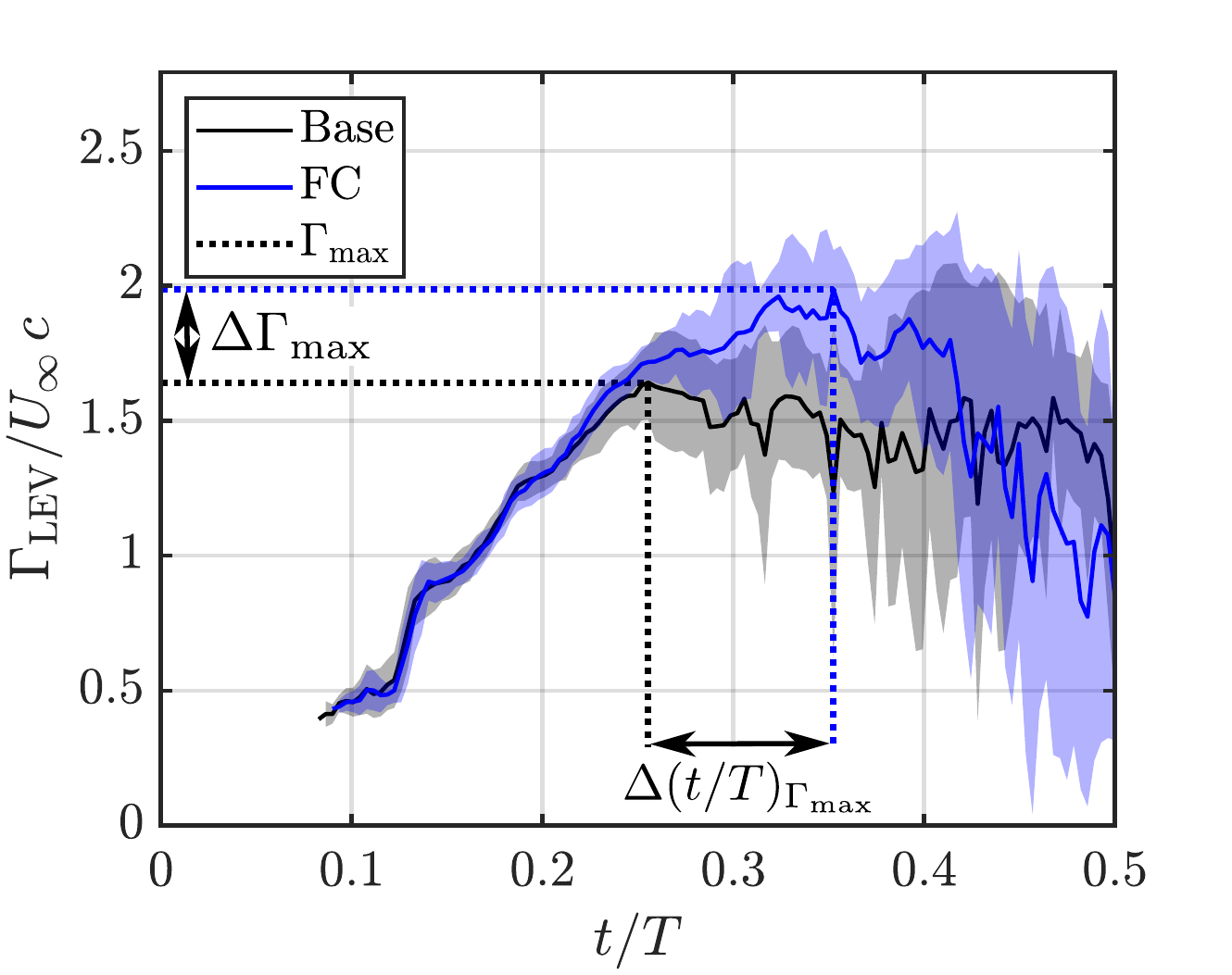}
    \label{fig:Gamma}
\end{subfigure}
\begin{subfigure}[t]{0.475\textwidth}
    \caption{}
    \includegraphics[width =\textwidth, trim = 0 5 25 15, clip]{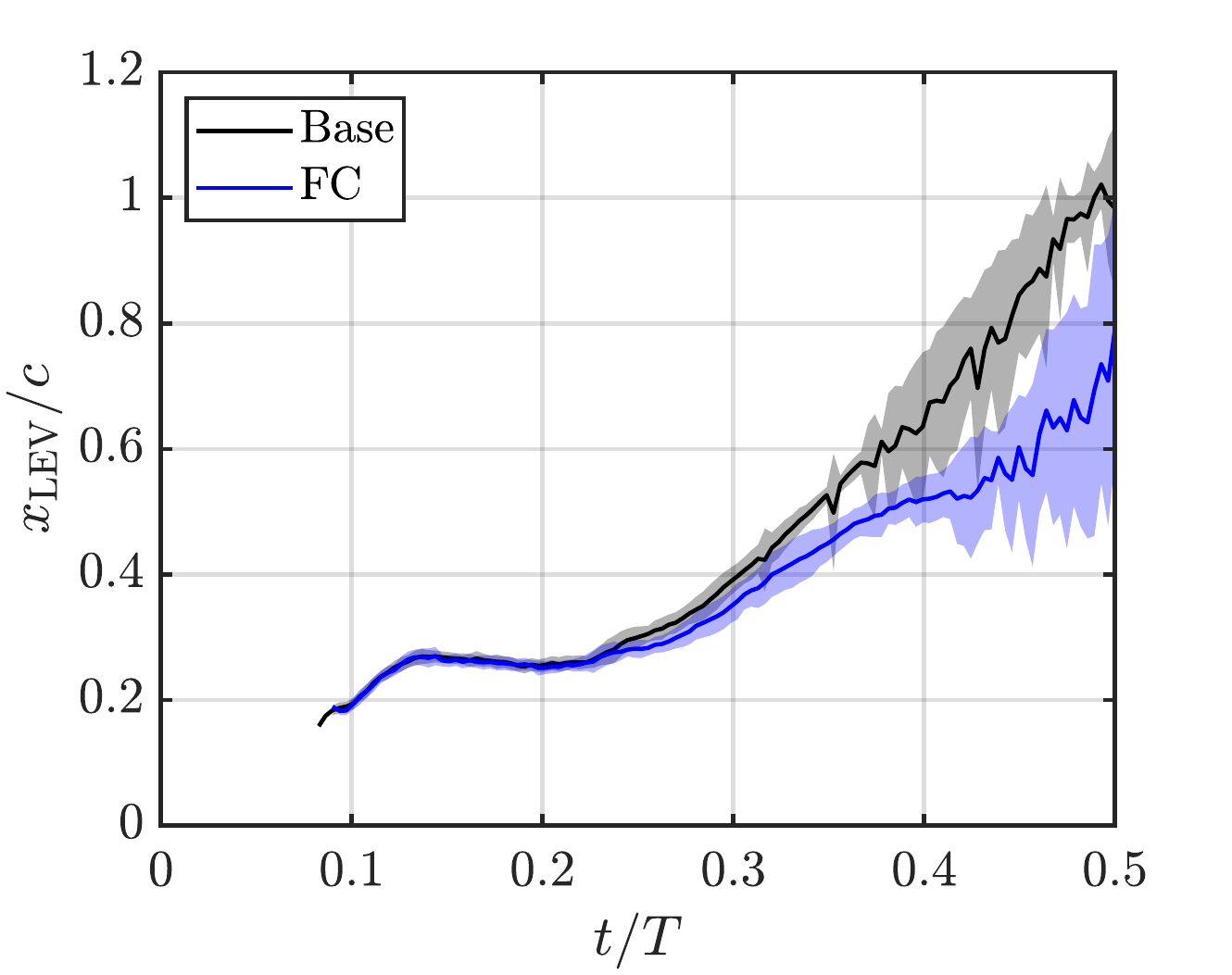}
    \label{fig:XPosition}
\end{subfigure}
\caption{Leading edge vortex characteristics evolution over dimensionless time $t/T$. The case without flow manipulation (Base) is shown in black and the controlled case (FC) in blue. The standard deviation between single runs is indicated by shaded areas of the respective color. (a) Normalized leading edge vortex circulation $\Gamma_{LEV}/U_{\infty} c$. The peak circulation $\Gamma_{\mathrm{max}}$ of the uncontrolled and controlled case is highlighted by dashed lines of the respective color in addition to the derived peak circulation increase $\Delta \Gamma_{\mathrm{max}}$ and temporal delay of peak circulation $\Delta (t/T)_{\Gamma_{\mathrm{max}}}$. (b) Normalized LEV center position $x_{\mathrm{LEV}}/c$}
\end{figure}

\textit{Manipulation effects on vortex characteristics.} To study the effect of secondary structure compression on the LEV detachment process, the LEV circulation and position is characterized, since a successful prolongation of the LEV growth phase, and thus higher induced lift, would be indicated by a prolonged circulation accumulation and a reduced center convection of the vortex.\\
The vortex circulation, obtained by integration of vorticity according to Stokes theorem within the detected vortex boundary obtained from $\Gamma_{2}$ scalar fields, is shown as a function of dimensionless time $t/T$ in Fig.~\ref{fig:Gamma}. 
Vortex identification results for all investigated cases are also available as raw data online (\url{http://dx.doi.org/10.25534/tudatalib-175}).
In the uncontrolled case (black curve), the LEV accumulates circulation up to $t/T\approx$ 0.25 and reaches a normalized peak circulation of about 1.6. In contrast, the controlled case (blue curve) leads to a LEV normalized peak  circulation of about 2. This is equivalent to a relative peak circulation increase $\Delta \Gamma_{\mathrm{max}}$ of 21\% with respect to the case without flow control.\\
The peak circulation in the controlled case is reached at about $t/T\approx$ 0.35. So the delay of peak circulation in the controlled case $\Delta (t/T)_{\Gamma_{\mathrm{max}}}$ is about 20\% with respect to the downstroke.\\
Note that both circulation histories collapse well beyond the start of the plasma actuator at $t/T=$ 0.15. In addition to  good repeatability of the flow scenario, this result also demonstrates that secondary structures, which are compressed in the controlled case, affect the vortex circulation significantly from $t/T=$ 0.25 onward.\\
A higher peak circulation of the LEV, which occurs later during the downstroke, indicates a prolonged LEV growth phase and thus, higher induced lift.\\
Another indicator for a prolonged growth phase of the LEV on the airfoil is its center convection. A reduced center convection indicates a delayed detachment of the vortex from the airfoil and thus, a longer phase of vortex induced lift. The normalized LEV center position $x_{\mathrm{LEV}}/c$ with and without flow control is compared in a plate-fixed frame of reference over $t/T$ in Fig.~\ref{fig:XPosition}. To extract the vortex center, the maximum of the $\Gamma_{1}$ scalar field was identified.\\
For the case where flow control is deployed (blue curve), the LEV center is located closer to the leading edge compared to the case without flow manipulation (black curve) from about $t/T=$ 0.25 onward. At the end of the downstroke, the LEV center of the controlled case is located 20\% more upstream compared to the case where flow control is not used.\\
In accordance with findings regarding the LEV circulation, its center convection is reduced by flow control and   infers a prolonged growth phase of the vortex.

\begin{figure}
    \begin{subfigure}[t]{0.54\textwidth}
        \includegraphics[width=\textwidth, trim = 0 0 0 0, clip]{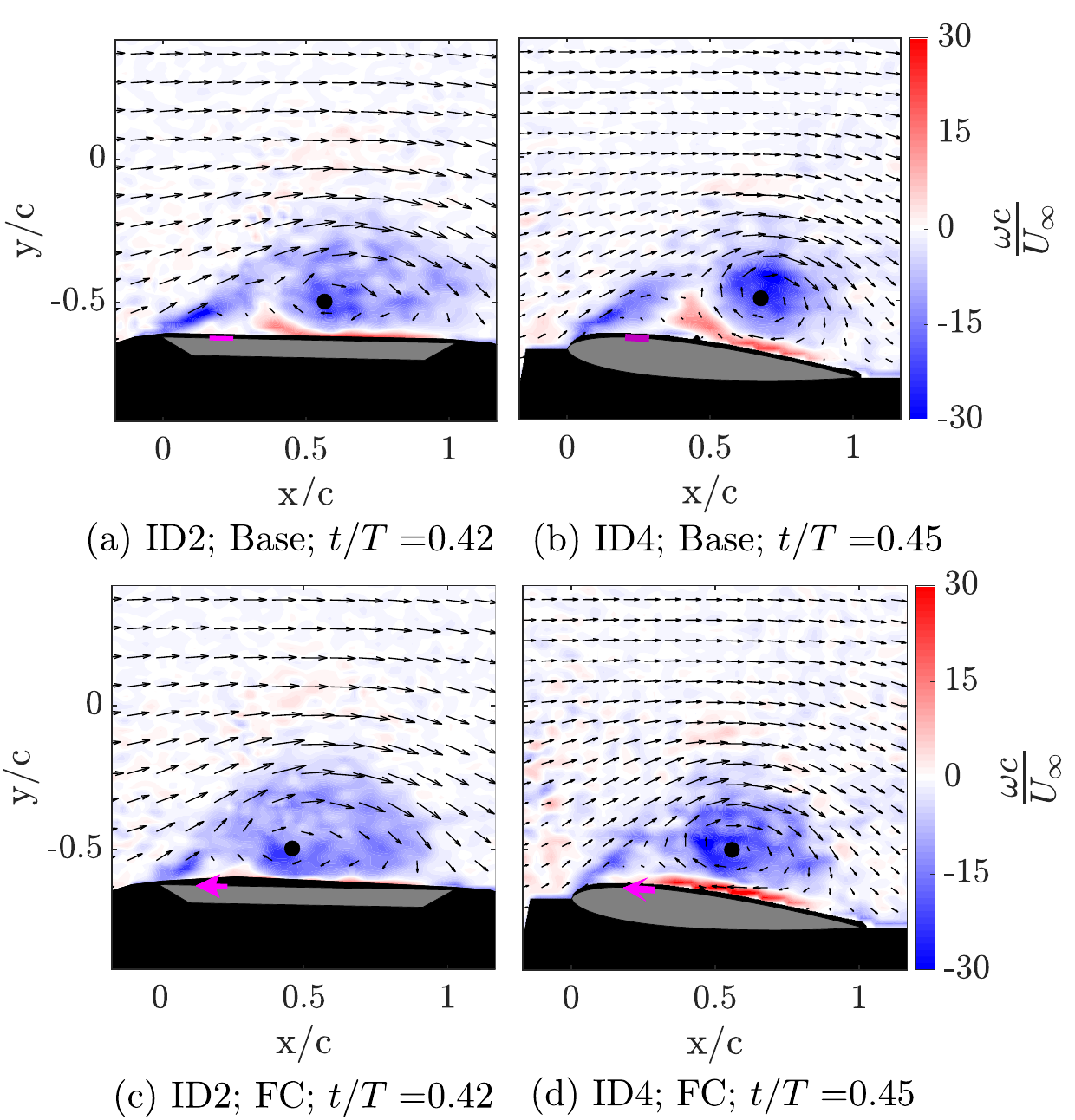}
    \end{subfigure}
    \begin{subfigure}[t]{0.44\textwidth}
        \includegraphics[width=\textwidth, trim = 0 0 0 0, clip]{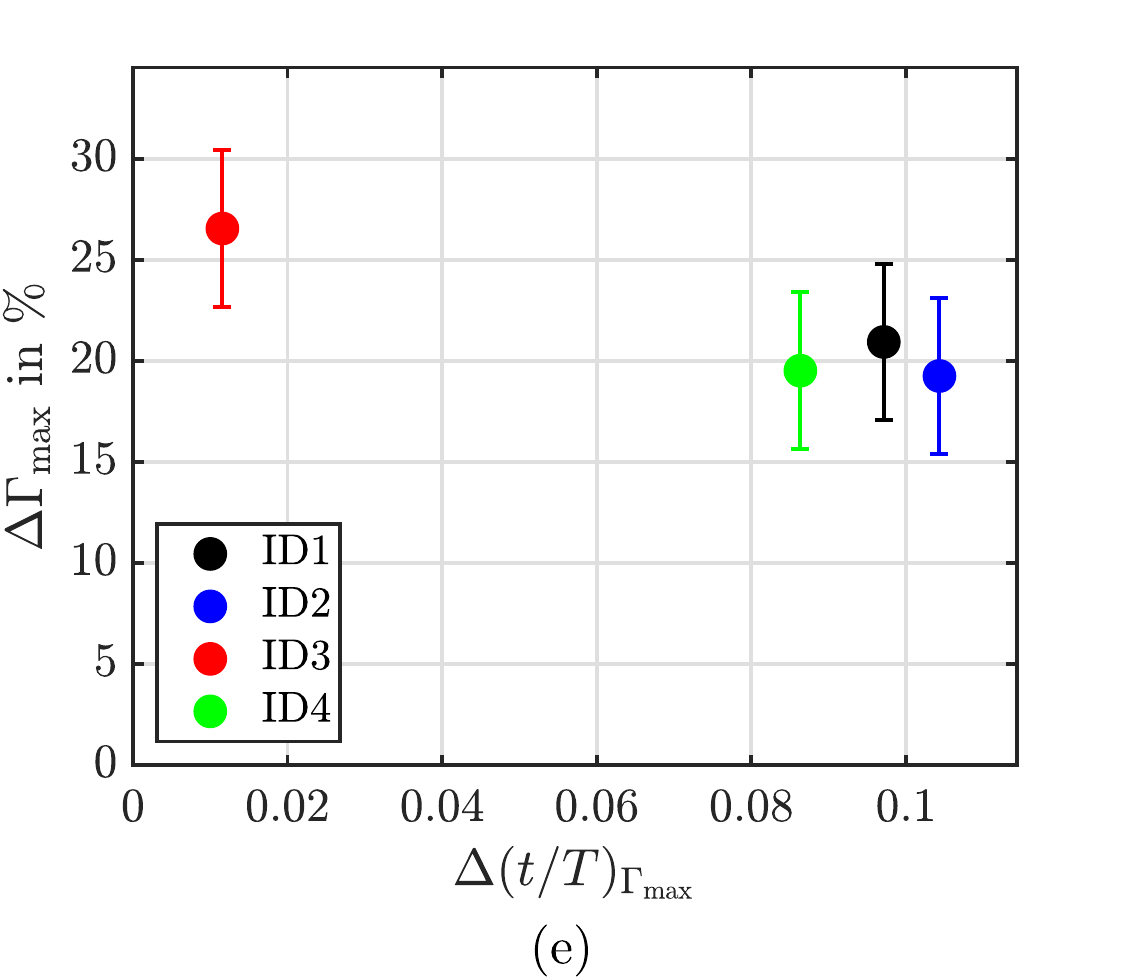}
    \end{subfigure}
\caption{\label{fig:OtherCases} (a)-(d) Unactuated (Base in the top row) and actuated (FC in the bottom row) flow field in terms of normalized vorticity $\omega c/ U_{\infty}$ for ID2 ((a) and (c)) and ID4 ((b) and (d)) cases at different dimensionless time instants $t/T$. (e) Relative peak circulation increase $\Delta \Gamma_{\mathrm{max}}$ between controlled and uncontrolled cases and the temporal delay of peak circulation $\Delta (t/T)_{\Gamma_{\mathrm{max}}}$. Error bars indicate the standard deviation of circulation increase.}
\end{figure}

\textit{Transferability of the manipulation hypothesis.} To test whether the introduced manipulation approach can be used to prolong the LEV growth phase also for other motion kinematics and on other airfoils, further flow control cases were tested, as listed in table~\ref{tab:table1}.
\begin{table}[b]
\caption{\label{tab:table1}%
Further test cases for plasma flow control according to the introduced manipulation approach. All cases at $Re$ = 22,000.
}
\begin{ruledtabular}
\begin{tabular}{cccddcdc}
\textrm{ID}&
\textrm{Sym.}&
\textrm{Airfoil}&
\textrm{$k$}&
\textrm{$St$}&
\textrm{$\alpha_{\mathrm{eff}}$ in\,$^{\circ}$}&
\textrm{$(x/c)_{\mathrm{PA}}$}&
\textrm{$\Delta(t/T)_{\mathrm{PA}}$}\\
\colrule
1 & \includegraphics[scale=0.4, trim = 0 0 0 0, clip]{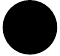} & Flat plate & 0.48 & 0.1 & 30 & 0.25 & 0.15-0.5\\
2 & \includegraphics[scale=0.4, trim = 0 0 0 0, clip]{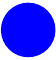} & Flat plate & 0.48 & 0.1 & 20 & 0.25 & 0.15-0.5\\
3 & \includegraphics[scale=0.4, trim = 0 0 0 0, clip]{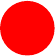} & Flat plate & 0.3 & 0.04 & 20 & 0.25 & 0.15-0.5\\
4 & \includegraphics[scale=0.4, trim = 0 0 0 0, clip]{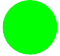} & NACA 0012 & 0.48 & 0.1 & 30 & 0.285 & 0.25-0.5\\
\end{tabular}
\end{ruledtabular}
\end{table}
ID1 represents the baseline case, discussed above. The ID2 case considers nearly pure plunging motion at an effective angle of attack $\hat{\alpha}_{\mathrm{eff}}$ of 20$\,^{\circ}$, resulting in a maximum geometric pitching angle of about 2$\,^{\circ}$. A lower Strouhal number and reduced frequency with respect to the ID2 case is investigated in the ID3 case. LEV manipulation on a NACA 0012 airfoil is tested in ID4, for which dimensionless and geometric parameters are kept the same as in ID1.\\
The plasma actuator location on the airfoil $(x/c)_{\mathrm{PA}}$ and the actuation period, where the plasma body force is deployed $\Delta(t/T)_{\mathrm{PA}}$, are determined by the procedure and criteria discussed above.\\
Exemplary flow fields with and without flow control for the ID2 and ID4 case at different dimensionless time instants $t/T$ are compared in Fig.~\ref{fig:OtherCases} (a)-(d).
In the ID2 case secondary structures can be identified by the positive (red) vorticity  upstream of the main leading edge vortex in Fig.~\ref{fig:OtherCases} (a). For the equivalent time instant in the controlled case, depicted in Fig.~\ref{fig:OtherCases} (c), no secondary structures can be identified.\\ 
This trend is even more pronounced for the ID4 case of the NACA 0012 airfoil (see Fig.~\ref{fig:OtherCases} (b) and (d)), which already indicates that the proposed manipulation hypothesis holds across a range of parameter combinations. Videos of the controlled and uncontrolled flow fields are also available online for all investigated cases (\url{http://dx.doi.org/10.25534/tudatalib-175}).\\
The relative LEV peak circulation increase $\Delta \Gamma_{\mathrm{max}}$ and the temporal delay of peak circulation $\Delta (t/T)_{\Gamma_{\mathrm{max}}}$ with and without flow control are used to evaluate the effectiveness of flow manipulation in terms of effects on vortex characteristics. Both measures indicate the termination of circulation accumulation of the LEV and thus allows  determination, whether or not the growth phase of the vortex is prolonged by flow control. They are exemplary indicated for the LEV circulation evolution in the ID1 case in Fig.~\ref{fig:Gamma} and shown for all cases in Fig.~\ref{fig:OtherCases} (e).\\
The peak circulation of the LEV increases between 19\,\% and 26\,\% when flow control is deployed. The peak circulation instant is delayed by about 20\,\% for ID1-ID2 and ID4 cases with respect to the downstroke, indicated by $\Delta (t/T)_{\Gamma_{\mathrm{max}}}$ values around 0.1.\\
For the ID3 case, the delay of peak circulation is smaller compared to the others. For this case, the circulation evolution without flow control reaches a plateau at about $(t/T)\approx$ 0.2, before it drops after $(t/T)=$ 0.3, similar to the circulation evolution for the ID1 case, shown in Fig.~\ref{fig:Gamma}. At the end of the plateau, the peak circulation in the uncontrolled case occurs and leads to a smaller temporal delay between the uncontrolled and controlled case $\Delta (t/T)_{\Gamma_{\mathrm{max}}}$.\\
However, Fig.~\ref{fig:OtherCases} (e) saliently demonstrates that the tested flow manipulation leads to significantly prolonged LEV growth and increased peak circulation for all investigated parameter combinations.\\
This highlights that the introduced manipulation hypothesis is also effective for very different motion kinematics and can be used to prolong the LEV growth phase on more realistic airfoils, like a NACA 0012. Additionally, due to the prolonged phase of vortex induced lift on the airfoil, the overall lift is assumed to be increased by flow control.

\textit{Conclusions.} This study addresses flow control of the leading edge vortex on pitching and plunging airfoils by dielectric barrier discharge plasma actuators. The overall goal is to prolong the vortex growth phase and thereby increase the vortex induced lift on the airfoil.\\
A manipulation hypothesis is introduced  based on flow manipulation at topologically critical locations and which aims to delay the vortex detachment process. The hypothesis is experimentally validated on a flat plate airfoil in deep dynamic stall and extended to different motion kinematics and more realistic airfoils.\\
The compression of secondary structures upstream of the LEV leads to a prolongation of the circulation accumulation phase and a reduced center convection of the LEV. Both are indicators for a prolonged growth phase and thus higher net lift.\\
Similar trends, with a peak circulation increase between 19\,\% and 26\,\%, are also observed for quasi pure plunging motion and lower motion dynamics of the flat plate airfoil and on a NACA 0012 airfoil, which demonstrates universal applicability of the manipulation approach.\\
A careful consideration of the actuation location and timing, in light of topological flow field developments identified by appropriate measures, is found to be crucial for flow control success.\\
The manipulation of the LEV detachment by a zero net mass flux device at topologically critical locations indicates the potential for an increased flow control efficiency with a minimum power input, based on a maximized lift enhancement.\\
Finally, higher efficiency of MAVs in cruise flight and hovering as well as an increased maneuverability and gust tolerance may be possible by using the proposed manipulation approach, which exploits beneficial lift effects of the LEV.

\textit{Acknowledgments.} The authors wish to acknowledge financial support of  the Sino-German Center and the Deutsche Forschungsgemeinschaft through the project TR 194/55-1: ``Flow Control for Unsteady Aerodynamics of Pitching/Plunging Airfoils". The authors would also like to extend their appreciation to the workshop staff in Darmstadt for their assistance and professional support for these experiments.

\bibliography{apssamp}

\end{document}